\documentclass[aps,twocolumn,showpacs,floats,prl]{revtex4}
\usepackage{graphicx}
\usepackage{dcolumn}
\usepackage{bm}
\usepackage{color}
\usepackage{amsmath}
\usepackage{subfigure}

\begin{document}

\author{L. A. Pe\~{n}a Ardila}
\author{W. Herrera}
\author{Virgilio ni\~{n}o}
\affiliation{Departamento de f\'{i}sica, Universidad Nacional de Colombia, Bogot\'{a}, Colombia}

\title{Momentum-energy tensor associated to the quasiparticles in  anisotropic superconductors} 

\begin{abstract} 
From a Lagrangian density for the Bogoliubov de Gennes equations in anisotropic superconductors, we find the momentum-energy tensor associated to the quasiparticles of the system. For this, we make infinitesimal translations on both space and
time and we use the Noether\rq{}s theorem. We prove that beyond to the usual terms associated to the electron-hole dynamic in electromagnetic potentials, appears  terms that involves the pair potential and that are obtained from the coupling of the electron-like and hole-like quasiparticles.  
\end{abstract}

\maketitle

\section{I. Introduction}
\label{sec:intro}

The Bogoliubov-de Gennes (BdG) equations describes the bahaviour of the elementary excitation of a superconductor system ~\cite{BGE,Leggett}.
Since there is  a Lagrangian density that allows one to derivate these equations and through of the Noether's theorem for infinitesimal transformations 
leaving the action invariant ~\cite{Goldstein}, it is possible to find the  conservation's laws associated to the superconductor system.
In this paper we build the momentum - energy tensor  from infinitesimal transformations both in space and time.

\section{Momentum-Energy tensor for anisotropic superconductors.}
\label{sec:tensor}
The Bogoliubov - de Gennes equations describe the quasiparticles in a superconductor are given by:
\begin{align}
\label{eq:Schrodinguer_Equation}
i\hbar\frac{\partial}{\partial t} \psi(\mathbf{r},t)=&\widehat{H}\psi(\mathbf{r},t)=\nonumber\\
\widehat{H}_{0}(\mathbf{r},t)\psi(\mathbf{r},t)+&\int dr \hat \Delta(\mathbf{r},\mathbf{r'})\psi(\mathbf{r'},t)
\end{align}

With,

\begin{align}
\label{eq:denside_prob_constante} 
\psi(\mathbf{r},t)=&\left(\begin{array}{cc} u(\mathbf{r},t)   \\ v(\mathbf{r},t)  \end{array}\right), \quad \widehat{H}_{0}(\mathbf{r},t)=\frac{\mathbf{ \hat \pi}^{2}(\mathbf{r},t)}{2m} \hat \sigma_{z} + \hat V(\mathbf{r},t) \nonumber\\
\hat V(\mathbf{r},t)=&(U(\mathbf{r},t)-\mu) \sigma_{z},\quad \hat \pi(\mathbf{r},t)=\left(\begin{array}{cc} \hat \pi_{e}(\mathbf{r},t) & 0 \\ 0  & \hat \pi_{h}(\mathbf{r},t) \end{array} \right) \nonumber\\
 \hat \Delta (\mathbf{r},\mathbf{r'})=&\left( \begin{array}{cc} 0 & \Delta (\mathbf{r},\mathbf{r'}) \\ \Delta^{*} (\mathbf{r},\mathbf{r'})  & 0 \end{array} \right), \quad \hat \sigma_{z} = \left( \begin{array}{cc} 1  & 0 \\ 0  & -1 \end{array} \right)
\end{align}

 where $\psi(\mathbf{r},t)$ is the wavefunction of the quasiparticle with two componets electron $u (\mathbf{r},t)$ and hole $v (\mathbf{r},t)$, whereas  $\hat \pi_{e}(\mathbf{r},t)=(-i\hbar\nabla-e\mathbf{\hat A})$ and $\hat \pi_{h}(\mathbf{r},t)=(-i\hbar\nabla+e\mathbf{\hat A})$ depictes the electron and hole momentum respectively; $ \mathbf{ \hat A}(\mathbf{r},t)$ is the magnetic vector potential,  $U(\mathbf{r},t)$ is the scalar potencial that includes both the Hartree-Fock and the external potencial, $\mu$ is the chemical potencial and   $ \hat \Delta (\mathbf{r},\mathbf{r'})$ is the pair potencial associated to quasiparticles and it is coupled to the electron and hole components. From the lagrangian density,

\begin{align}
\label{eq:Lagrangiandensity}
\mathcal{L}=&\frac{i \hbar}{2}[\psi^{+}(\mathbf{r},t)\dot \psi(\mathbf{r},t)-\dot\psi^{+}(\mathbf{r},t)\psi(\mathbf{r},t)] \nonumber\\
&-\psi^{+}(\mathbf{r},t) \hat V(\mathbf{r},t) \hat \sigma_{z}\psi(\mathbf{r},t)\nonumber\\
&-\frac{1}{2m}[\hat P(\mathbf{r},t)\psi((\mathbf{r},t)]^{+} \sigma_{z}\hat P(\mathbf{r},t)\psi(\mathbf{r},t))^{+}\nonumber\\
&-\int dr^{'} \psi(\mathbf{r},t)^{+} \Delta (\mathbf{r},\mathbf{r'}) \psi(\mathbf{r'},t).
\end{align}

The Bogoliubov - de Gennes (BdG) equations are derived by means of the principle of least action ~\cite{Sym}. The Noether theorem allows us to find both a set of equations and conservation laws from transformations that leave the action invariant. If  an infinitesimal transformation of $p$ parameters that leaves action unchangeable, then there are $p$ quantities conserved ~\cite{Hill}. If the motion equation for the fields $\Phi_{k}$ is obtained from the principle of least action and moreover the transformations in space, time and the fields leaves the action invariant. The Noether theorem yields:

\begin{equation}
\label{eq:Conservation}
\frac{\partial\rho}{\partial t} +\nabla \cdot \mathbf{J} = 0,
\end{equation}

\begin{equation}
\label{eq:rho}
\rho=\mathcal{L}\delta t - \frac{\partial\mathcal{L}}{\partial (\dot \Phi_{k})} \left(\sum_{j=1}^{3} \partial_{j}\Phi_{k}\delta x_{j}+ \dot \Phi_{k}\delta t- \delta \Phi_{k}\right)
\end{equation}

\begin{equation}
\label{eq:Current}
J_{i}=\mathcal{L}\delta x_{i} - \frac{\partial\mathcal{L}}{\partial (\partial_{i} \Phi_{k})}\left(\sum_{j=1}^{3} \partial_{j}\Phi_{k}\delta x_{j}+ \dot \Phi_{k}\delta t- \delta \Phi_{k}\right)
\end{equation}

Where $\rho$ and $\mathbf{J}$ correspond to the charge and current densities associated to the infinitesimal transformation. Moreover when  Eq. [\ref{eq:Current}] is integrated over all space, one obtains:
\begin{equation}
\label{eq:Current2}
\frac{d}{dt} \int d^{3} \mathbf{r} \rho(\mathbf{r},t) = \frac{dQ(\mathbf{r},t)}{dt}=0.
\end{equation}
Where Q is a generalized charge and is itself  the conserved quantity.

\section{Temporal translation-conservation of the energy}
\label{sec:Energy}

Let us do an infinitesimal transformation in time, in this way:

\begin{equation}
\label{eq:time_translation}
t^{'}=t+\delta t.
\end{equation}

The variations respect to the fields and spatial coordinates are vanished.By plugging the lagrangian density Eq. [\ref{eq:Lagrangiandensity}] into Eq. [(\ref{eq:rho})], we obtain:

\begin{align}
\label{eq:p}
\rho_{0}(\mathbf{r},t)=& \frac{1}{2m}[\hat \pi(\mathbf{r})\psi(\mathbf{r},t)]^{+} \sigma_{z} \cdot \pi(\mathbf{r})\psi(\mathbf{r},t)\nonumber\\ 
+&\psi^{+}(\mathbf{r},t) \hat V(\mathbf{r}) \psi(\mathbf{r},t) + \int dr \psi^{+}(\mathbf{r},t) \Delta(\mathbf{r},\mathbf{r'}) \psi(\mathbf{r},t) 
\end{align}

This expression can be interpreted as an energy density that is shown as the sum of the energy density associated with the electron and hole components plus the energy density that couples electrons and holesdepending of the pairs potential $\Delta$.

By using  Eq. [\ref{eq:Current2}] and recalling that this transformation leaves the action invariant, we have $\frac{d}{dt}\int \rho_{0} dV=\frac{d}{dt}\left<E \right>=0$, which implies that the expectation value of energy is a constant of motion. Now let us examine what happens with the density of current in this case. From Eq. [\ref{eq:rho}] for the current density and again using  the lagrangian density Eq. [ \ref{eq:Lagrangiandensity}], we have

\begin{align}
\label{eq:jo}
J_{0,j}=&\frac{1}{2}\left[(V_{j}\psi)^{+}\widehat{H}\psi + (\widehat{H}\psi)^{+}(V_{j}\psi)\right]\nonumber \\ 
&=\frac{1}{2} \left[(-i\hbar \hat \sigma_{z}\partial_{j}-eA_{j}  \hat 1) \psi \right]^{+} \widehat{H}\psi \nonumber \\ 
&+ (\widehat{H}\psi)^{+})(-i\hbar \hat \sigma_{z}\partial_{j}-eA_{j}  \hat 1) \psi \quad j=1,2,3.
\end{align}
Where $\frac{\hat \sigma_{z}\hat \pi_{z}}{m} = \mathbf{\hat v}$ is the velocity of group of the particle. From this equation is observed that $\mathbf{J}$ is an energy flux, because it can be  written as the quasiparticle  density of charge times its group velocity.

\section{Spatial  translation-Conservation of linear momentum}
\label{sec:Momentum}

Now we can do an infinitesimal translation in coordinates
\begin{equation}
\label{eq:temporal_translation}
x_{i}^{'}=x_{i}+\delta x_{i} \quad i=1,2,3.
\end{equation}

By using the Eq. [\ref{eq:rho}] for each tspatial component $x$, $y$, and $z$ we get a charge density associated   given by:
\begin{equation}
\label{eq:chargeDensity}
\rho_{i}=\psi^{+}\hat \pi_{i} \psi = \psi^{+} (-i \hbar \partial_{i} \hat 1- eA_{i} \hat \sigma_{z}) \psi
\end{equation}

In the same way. Using Eq. [\ref{eq:Current}] the current density associated to $\rho_{i}$ is given by: 

\begin{align}
\label{eq:J}
J_{i,j}=&\frac{1}{2}\left[(\hat V_{j} \psi)^{+} \hat \pi_{i} \psi-(\hat \pi_{i}\psi)^{+} \hat V_{j} \psi \right]\nonumber\\
=&\frac{1}{2}[[(-i \hbar \hat \sigma_{z} \partial_{j}-eA_{j} \hat 1)  \psi]^{+}(-i \hbar  \partial_{i}\hat1-eA_{i} \hat \sigma_{z})\psi\\
+&[(-i \hbar \partial_{i} \hat 1  -eA_{i} \hat \sigma_{z}) \psi]^{+}(-i \hbar \hat \sigma_{z} \partial_{j}-eA_{j} \hat 1) \psi ]\nonumber\\
\end{align}

We obtain that Eq. [\ref{eq:chargeDensity}] corresponds to the density of linear momentum, whereas Eq. [ \ref{eq:J}] corresponds to the density of current of momentum, that is proportional to the density of momentum and to the velocity of group of the quasiparticle. If the action is invariant under these translations. one can integrate over all space the density of momentum and it yields
\begin{align}
\label{eq:impondo_autoconsistencia}
\frac{d}{dt}\int \psi^{+} \hat \pi_{i} \psi dV = \frac{d}{dt} \left<\pi_{i} \right>
\end{align}

Therefore the momentum associated to the quasiparticles is conserved. The density of momentum can be written as $\rho_{i}=\rho_{ei}+\rho_{hi}$, where $\rho_{ei}$ and $\rho_{hi}$ are the momentum density associated to electron and hole components, respectively. So we have

\begin{align}
\label{eq:impondo_autoconsistencia}
\rho_{ei}=u^{*}(-i \hbar \partial_{i}-eA_{i})u \quad; \rho_{hi}=v^{*}(-i \hbar \partial_{i}+eA_{i})v.
\end{align}

In the case that action is not invariant under transformation Eq.[\ref{eq:chargeDensity}], there will appear sources proportional to the  scalar potential gradient, magnetic vector potencial and the pairs potential. All the equations we have derived so far  from the space-time transformations are summarized  in a tensorial form :


\[T_{uv}= \left( \begin{array}{cccc}
\rho_{0} & J_{01} & J_{02}  & J_{03} \\
\rho_{1} & J_{11} & J_{12}  & J_{13} \\
\rho_{2} & J_{21} & J_{22}  & J_{23} \\
\rho_{3} & J_{31} & J_{32}  & J_{33} \\  \end{array}\right)\]

Where the term $T_{00}$ corresponds to the density of energy of the quasiparticle, whose explicit form is given in  Eq. [\ref{eq:p}], the terms of the form $T_{0i}$ $i=1,2,3$ are related with the the density of current of energy in each component that have been calculated in Eq.[\ref{eq:jo}]. Now the terms of $T_{i0}$  represents the density of linear momentum in the direction i given by Eq. [\ref{eq:chargeDensity}] and finally the terms of the form  $T_{ij}$  corresponds to the component j of the momentum current density  is given by Eq. [\ref{eq:J}]. In conclusion, the tensorial form for the energy and momentum equations written as:

\begin{equation}
\label{eq:Tensor}
\frac{dT_{uv}}{dx_{u}}=0  \quad u,v=0,1,2,3.
\end{equation}

When a spatial or temporal infinitesimal transformation does not leave the  lagrangian density invariant an associated source appears. In classical mechanics, when a potential does not depend on the position the total force is zero and is a conserved quantity. In contrast, if the potential does depend of the position the quantity is not conserved and the source is the total force. In the next section we will see that the langrangian Eq. [\ref{eq:Lagrangiandensity}] is not invariant if for instance, it contains a potential vector dependence $\mathcal{L}(\mathbf{A}(x))$.

\section{Example: Lorentz force as a source from a lagrangian depending on the potential vector}

A particle with mass $m$ and electric charge $q=-e$ in presence
of a magnetic field can be described by:

\begin{equation}
{\hat{H}\psi(\mathbf{r},t)}=\frac{\left(\mathbf{\hat p}-e\mathbf{\hat A}\right)^{2}}{2m}\psi(\mathbf{r},t)=-i\hbar\frac{\partial\psi(\mathbf{r},t)}{\partial t}\label{eq:hamiltonian_magnetic_field}
\end{equation}

where $\mathbf{\hat p}=-i\hbar\mathbf{\nabla}$. A magnetic field is generated
by a potential vector $\mathbf{\hat A}$. From Eq. [\ref{eq:hamiltonian_magnetic_field}] one obtains the continuity equation Eq. [\ref{eq:Conservation}] with,
 
\begin{equation}
\hat{J}=\frac{1}{2m}\left[\psi^{+}\left(-i\hbar\nabla-q\mathbf{\hat A}\right)\psi-\psi\left(-i\hbar\nabla+q\mathbf{\hat A}\right)\psi^{+}\right]\label{eq:current}
\end{equation}

The temporal evolution of an operator $\hat{\mathrm{O}}$ is described
by
\begin{equation}
i\hbar\frac{d}{dt}\left\langle \hat{\mathrm{O}}\right\rangle =\left\langle \left[\hat{\mathrm{O}},\hat{H}\right]\right\rangle 
\end{equation}

By definition the conjugate momentum is written as, $\vec{\pi}=\mathbf{p}-e\mathbf{A}$
and the velocity is given by:

\begin{equation}
\overset{\cdot}{\mathbf{\hat x}}=\mathbf{\hat v}=\frac{\mathbf{\hat p}-e\mathbf{\hat A}}{m}=\frac{\vec{\hat \pi}}{m}=\left\langle \frac{\mathbf{\hat p}-e\mathbf{\hat A}}{m}\right\rangle 
\end{equation}

therefore $\hat{\pi}/m$ plays the role of velocity and the temporal
evolution of $\hat {\pi}$ will give rise to the quantum
Lorentz force, rather than the temporal evolution of $\mathbf{\hat p}$.
In fact,

\begin{equation}
\frac{d}{dt}\left\langle \hat p_{i}\right\rangle =\frac{q}{2}\left\langle \partial_{i}\hat A^{j}\hat v_{j}+\hat v_{j}\partial_{i}\hat A^{j}\right\rangle \label{eq:momentum_evolution}
\end{equation}

Since $\mathbf{\hat v}=\frac{\mathbf{\hat p}-e\mathbf{\hat A}}{m}$ clearly Eq. [\ref{eq:momentum_evolution}] does not match with the Lorentz force definition. Here repeated indexes
sum is considered and $i=1,2,3$ are the coordinates. Let us find
the temporal evolution of the potential vectorial $\mathbf{\hat A}$ :

\begin{equation}
i\hbar\frac{d}{dt}\left\langle \hat A^{i}\right\rangle =\left\langle \left[\hat A^{i},{\hat H}\right]\right\rangle 
\end{equation}
\begin{equation}
i\hbar\frac{d}{dt}\left\langle \hat A^{i}\right\rangle =\left\langle \left[\hat A^{i},\frac{\left(\hat p_{j}+q \hat A_{j}\right)^{2}}{2m}\right]\right\rangle 
\end{equation}

\begin{equation}
i\hbar\frac{d}{dt}\left\langle \hat A^{i}\right\rangle =\left\langle \left[\hat A^{i},\frac{\left(\hat p_{j}+q\hat A_{j}\right)\cdot\left(\hat p_{j}+q\hat A_{j}\right)}{2m}\right]\right\rangle 
\end{equation}

we can use the fact that $[\hat{a},\hat{b}^{2}]=[\hat{a},\hat{b}]\hat{b}+\hat{b}[\hat{a},\hat{b}].$
Therefore, straightforwardly

\begin{equation}
\begin{array}{c}
i\hbar\frac{d}{dt}\left\langle \hat A^{i}\right\rangle =\frac{1}{2m}\left\langle \left[\hat A^{i},\left(\hat p_{j}+q \hat A_{j}\right)\right]\left(\hat p_{j}+q\hat A_{j}\right)\right.\\
\left.+\left(\hat p_{j}+q\hat A_{j}\right)\left[\hat A^{i},\left(\hat p_{j}+q\hat A_{j}\right)\right]\right\rangle 
\end{array}
\end{equation}

Again ones associates $\left(\hat p_{j}+q\hat A_{j}\right)=m\hat v_{j}=\hat \pi_{j}$. Therefore

\begin{equation}
i\hbar\frac{d}{dt}\left\langle \hat A^{i}\right\rangle =\frac{1}{2}\left\langle \left[\hat A^{i},\left(\hat p_{j}+q\hat A_{j}\right)\right]\hat v_{j}+\hat v_{j}\left[\hat A^{i},\left(\hat p_{j}+q\hat A_{j}\right)\right]\right\rangle 
\end{equation}

Evidently $[\hat A^{i},\hat A^{j}]=0$. One ends up with:

\begin{equation}
i\hbar\frac{d}{dt}\left\langle \hat A^{i}\right\rangle =\frac{1}{2}\left\langle \left[\hat A^{i},\hat p_{j}\right]v_{j}+\hat v_{j}\left[\hat A^{i},\hat p_{j}\right]\right\rangle 
\end{equation}

Easily one shows that $[\hat A^{i},\hat p_{j}]=i\hbar\partial_{j}\hat A^{i}$. Then,

\begin{equation}
i\hbar\frac{d}{dt}\left\langle \hat A^{i}\right\rangle =\frac{i\hbar}{2}\left\langle \partial_{j} \hat A^{i}v_{j}+\hat v_{j}\partial_{j} \hat A^{i}\right\rangle \label{vector_potential_evolution}
\end{equation}

Results from either Eq. [\ref{eq:momentum_evolution}] and Eq. [\ref{vector_potential_evolution}] give us a direct way to calculate the temporal evolution of the conjugate momentum. It means,

\begin{equation}
\frac{d}{dt}\left\langle \hat \pi^{i}\right\rangle =\frac{d}{dt}\left\langle \hat p^{i}-e \hat A^{i}\right\rangle 
\end{equation}

Plugging results from Eq. [\ref{eq:momentum_evolution}] and Eq. [\ref{vector_potential_evolution}] yields,

\begin{equation}
\frac{d}{dt}\left\langle \hat \pi^{i}\right\rangle =\frac{q}{2}\left\langle \left(\partial_{i} \hat A^{j}-\partial_{j} \hat A^{i}\right) \hat v_{j}+ \hat v_{j}\left(\partial_{i} \hat A^{j}-\partial_{j} \hat A^{i}\right)\right\rangle 
\end{equation}

rewritting in terms of the magnetic field, 

\begin{equation}
\frac{d}{dt}\left\langle \hat \pi^{i}\right\rangle =\frac{q}{2}\left\langle \epsilon^{ijk}\left(\hat B^{k}\hat v_{j}+\hat v_{j}\hat B^{k}\right)\right\rangle 
\end{equation}

That yields directly the definition of the Lorentz force:

\begin{equation}
\frac{d}{dt}\left\langle \hat{\pi}\right\rangle =\frac{q}{2}\left\langle \mathbf{\hat v}\times\mathbf{\hat B-\mathbf{\hat B\mathbf{\times \hat  v}}}\right\rangle 
\end{equation}

it has been showed that the Lorentz force is derived from the canonical
momentum $\hat \pi$ rather than its conjugate one $\mathbf{\hat p}$.

The current equation Eq. [\ref{eq:current}] can be written is terms of velocity as:

\begin{equation}
\mathbf{J}=\frac{1}{2m}\left[\psi^{+}\left(-i\hbar\nabla-q\mathbf{\hat A}\right)\psi-\psi\left(-i\hbar\nabla+q\mathbf{\hat A}\right)\psi^{+}\right]
\end{equation}

\begin{equation}
\mathbf{J}=\frac{1}{2m}\left[\psi^{+}\mathbf{\hat v}\psi-\left(\mathbf{\hat v}\psi\right)^{+}\psi\right]\label{eq:J}
\end{equation}

and from Eq. [\ref{eq:momentum_evolution}],

\begin{equation}
\begin{array}{c}
\frac{d}{dt}\left\langle \hat p_{i}\right\rangle =\frac{q}{2}\left\langle \partial_{i}\hat A^{j} \hat v_{j}+\hat v_{j}\partial_{i} \hat A^{j}\right\rangle =\\
\int d\mathbf{r}\psi^{+}\left(\partial_{i}\hat A^{j}\hat v_{j}+\hat v_{j}\partial_{i}\hat A^{j}\right)\psi\\
=\int d\mathbf{r}\left[\psi^{+}\hat v_{j}\psi+\left(\hat v_{j}\psi^{+}\right)\psi\right]\partial_{i}\hat A^{j}=\int d\mathbf{r}J^{j}\partial_{i}\hat A^{j}
\end{array}
\end{equation}

Analogously, the lagrangian density can be written as:

\begin{equation}
\mathcal{L}=\frac{i\hbar}{2}\left[\psi^{+}\partial_{i}\psi-\partial_{i}\psi^{+}\psi\right]-\hat{H}
\end{equation}

and

\begin{equation}
\mathcal{L}_{0}=\frac{i\hbar}{2}\left[\psi^{+}\partial_{i}\psi-\partial_{i}\psi^{+}\psi\right]-\mathcal{\left(\hat \pi\psi\right)}^{+}\left(\hat \pi\psi\right)
\end{equation}

where the hamiltonian density is defined as ${\hat H=\left(\pi\psi\right)}^{*}\left(\pi\psi\right)$
and $\mathcal{L}_{0}$ is the free lagrangian density (lagrangian
density with $\mathbf{\hat A}=0.$)

The lagrangian density previously written is split in this convinient
way:

\begin{equation}
\mathcal{L}=\mathcal{L}_{0}+\mathbf{J\cdot \hat A}\label{eq:l-lo}
\end{equation}

and $\mathbf{J}$ is given by Eq. [\ref{eq:J}]. Let us considera couple of cases of interest: whether or not the vectorial potential
has a dependence with position.

\subsection{Constant vectorial potential }

In this case, one has that the langrangian is invariant under spacial translation and the associated charged is conserved. In other words

\begin{equation}
\partial_{i}T_{\nu}^{\upsilon}=0
\end{equation}

where the energy-momentum tensor is written explicity as 

\begin{equation}
T_{\nu}^{\upsilon}=\frac{\delta\mathcal{L}}{\delta\partial_{\mu}\phi_{\alpha}}\partial_{\upsilon}\phi_{\alpha}-\mathcal{L}\phi_{\alpha}-\delta\mathcal{L_{\mu}^{\upsilon}}
\end{equation}

Recalling for the momentum-component:

\begin{equation}
T_{i}^{0}=\frac{1}{2}i\hbar\psi^{+}\partial_{i}\psi-i\hbar\partial_{i}\psi^{+}\psi
\end{equation}

\subsection{Position-dependence of the vector potential.}

by performing an spacial translation along the $i$ dirrection, it
yields that

\begin{equation}
\int d^{3}\mathbf{x}\left[\mathcal{L}(x^{i}+\delta x^{i})-\mathcal{L}(x^{i})\right]=-\int d^{3}\mathbf{x}\partial_{u}T_{i}^{u}\delta x_{i}
\end{equation}

The vector potential is non invariant under spatial translation. It
gives:

\begin{equation}
\left[\mathcal{L}(x^{i}+\delta x^{i})-\mathcal{L}(x^{i})\right]=\frac{\delta\mathcal{L}}{\delta \hat A^{j}}\partial_{i}\hat A^{j}\delta x_{i}
\end{equation}

From Eq. {[}\ref{eq:l-lo}{]} one knows that $\delta\mathcal{L}/\delta \hat A^{j}=qJ^{i}$,
therefore 

\begin{equation}
\int d^{3}\mathbf{x}\left[\mathcal{L}(x^{i}+\delta x^{i})-\mathcal{L}(x^{i})\right]=-\int d^{3}\mathbf{x}\partial_{u}T_{i}^{u}
\end{equation}

\begin{equation}
\begin{array}{c}
\int d^{3}\mathbf{x}qJ^{i}\partial_{i}\hat A^{j}=-\int d^{3}\mathbf{x}\partial_{u}T_{i}^{u}\\
=-\int d^{3}\mathbf{x}\partial_{0}T_{i}^{0}\delta x_{i}=\frac{d\left\langle \hat p^{i}\right\rangle }{dt}
\end{array}
\end{equation}

In conclusion, 

\begin{equation}
\frac{d}{dt}\left\langle \hat \pi\right\rangle =\frac{d}{dt}\left\langle \mathbf{\hat p}-e\mathbf{\hat A}\right\rangle =\frac{d}{dt}\left\langle \mathbf{\hat p}\right\rangle -e\frac{d}{dt}\left\langle \mathbf{\hat A}\right\rangle 
\end{equation}

\begin{equation}
\frac{d}{dt}\left\langle \mathbf{\hat p}\right\rangle =\frac{q}{2}\left\langle \mathbf{\hat v}\times\mathbf{\hat B-\mathbf{\hat B\mathbf{\times \hat v}}}\right\rangle -e\frac{d}{dt}\left\langle \mathbf{\hat A}\right\rangle 
\end{equation}

And finally:

\begin{equation}
\frac{d}{dt}\left\langle \hat p^{i}\right\rangle =\hat F_{Lorentz}^{i}-\frac{e}{2}\left\langle \partial_{j}\hat A^{i}v_{j}+\hat v_{j}\partial_{j}\hat A^{i}\right\rangle 
\end{equation}

\section{Conclusions}\label{sec:conclusion}

From infinitesimal transformations both in space and time  we have found the Energy-Momentum tensor  associated to quasiparticles for an anisotropic superconductor. In the case  where the transformations leave the action invariant , we find that the energy and linear momentum of quasiparticles are the quantities conserved.

\section*{Acknowledgments}
The author thanks to the Research divisi\'{o}n of the Universidad Nacional De Colombia for the financial support.


\begin{thebibliography}{99}
\bibitem{BGE} P.G. De Gennes, Superconductivity of Metals and Alloys (Benjamin, New York, 1966).
\bibitem{Leggett} Quantum Liquids,Bose Condensation and Cooper Pairing in Condensed-Matter Systems, Anthony James Leggett, Oxford Graduate Texts,2006.

\bibitem{Goldstein}Goldstein, Herbert (1980). Classical Mechanics (2nd ed.). Reading, MA: Addison-Wesley. pp. 588–596.

\bibitem{Sym} W. Greiner, Quantum mechanics Symmetries 2a, (1995).
\bibitem{Hill} Rev. Mod. Phys. 23, 253,1951
\end{thebibliography}
\end{document}